# Cooperative control of perpendicular magnetic anisotropy via crystal structure and orientation in single-crystal flexible SrRuO$_3$ membranes

*Zengxing Lu, Yongjie Yang, Lijie Wen, Jiatai Feng, Bing Lao, Xuan Zheng, Sheng Li, Kenan Zhao, Bingshan Cao, Zeliang Ren, Dongsheng Song, Haifeng Du, Yuanyuan Guo, Zhicheng Zhong, Xianfeng Hao,[\*] Zhiming Wang,[\*] and Run-Wei Li*

**ABSTRACT:** Flexible magnetic materials with robust and controllable perpendicular magnetic anisotropy (PMA) are highly desirable for developing flexible high-performance spintronic devices. However, it is still challenge to fabricate PMA films through current techniques of direct deposition on polymers. Here, we report a facile method for synthesizing single-crystal freestanding SrRuO$_3$ (SRO) membranes with controlled crystal structure and orientation using water-soluble Ca$_{3-x}$Sr$_x$Al$_2$O$_6$ sacrificial layers. Through cooperative effect of crystal structure and orientation engineering, flexible SrRuO$_3$ membranes reveal highly tunable magnetic anisotropy from in-plane to our-of-plane with a remarkable PMA energy of 7.34×10$^6$ erg/cm$^3$. Based on the first-principles calculations, it reveals that the underlying mechanism of PMA modulation is intimately correlated with structure-controlled Ru 4d-orbital occupation, as well as the spin-orbital matrix element differences, dependent on the crystal orientation. In addition, there are no obvious changes of the magnetism after 10,000 bending cycles, indicating an excellent magnetism reliability in the prepared films. This work provides a feasible approach to prepare the flexible oxide films with strong and controllable PMA.

**INTRODUCTION**

Flexible electronic/spintronic devices, with bendable and lightweight features, greatly expand the boundaries of spintronic/electronic applications and are gradually changing personal habits in our daily life.[1-6] As one of core researching goals of the flexible spintronics/electronics, developing high-performance devices is highly coveted.[4-6] In this regard, integrating perpendicular magnetic anisotropy (PMA) to flexible devices shows tremendous potential. The PMA shows advantage in enhancing device density and lowering energy consumption while preserving superior thermal stability.[7-9] Moreover, controlling and tailoring the PMA on purpose plays a significant role in the formation of emergent spin textures such as chiral domain walls and magnetic skyrmions, which are considered as important ingredient for next-generation spintronic devices.[10,11] Currently, although a great number of the flexible PMA materials have been synthesized directly on flexible bases[4,12-14], the limitations of the polymer substrates, such as low melting point and large roughness, hinder to obtain films with excellent quality and robust PMA using the direct deposition technique[15,16]. Hence, it is desirable to develop facile method for synthesis of high-quality flexible materials with a strong and controllable PMA.

$SrRuO_3$ (SRO) is an attractive candidate for realizing flexible PMA-based spintronic materials and devices. It possesses itinerant ferromagnetism and strong spin-orbit coupling simultaneously, resulting in an impressive magnetic crystalline anisotropy (MCA) with an anisotropy energy (MAE) above $10^6$ erg/cm$^3$.[17-20] Moreover, the SRO is a 4d transition metal oxide (TMO), which has complex interactions between multiple degrees of freedom, and can trigger multiple types of phase transformations controlled by various external stimuli.[21-24] These features offer an opportunity to realize the strong and controllable PMA in the freestanding SRO membranes. Nowadays, for synthesizing the flexible films, one of promising methods is to utilize sacrificial buffer layers through a combination of lift-off and thin-film transfer techniques.[25,26] Pioneered by Lu et al. in 2016[27], the water-soluble $Sr_3Al_2O_6$ (SAO) sacrificial layer has been widely used to synthesize a variety of freestanding crystalline TMOs membranes with millimeter-scale lateral size, controllable thickness down to ultrathin limit and extremely large tolerable strain.[28-36] Moreover, different crystalline oriented oxide membranes can be readily fabricated by changing the substrate crystal orientation, which provides a feasible way to control the properties of oxide membranes.[24, 35, 37, 38] In addition, by controlling the thickness and composition of sacrificial layer, it can further modulate the properties of freestanding TMOs membranes.[39, 40]

In this work, using the SAO and $Ca_{1.5}Sr_{1.5}Al_2O_6$ (CSAO) as sacrificial layers, we prepare freestanding SRO films with different growth orientations, and realize a strong and highly tunned PMA. Intriguingly, the crystal structures of flexible membranes are tetragonal and orthorhombic when deposited with the SAO and CSAO, respectively. The different structures have a great influence on the PMA and the effect becomes more prominent with the collaboration of the orientation. By altering the growth orientation of the films, firstly, the easy-axis of the tetragonal SRO is controlled. The [110]- and [111]-oriented films show in-plane MA (IMA) while the [001]-film owns a PMA. In contrary, for the orthorhombic SRO, the PMA is the protagonist in all of the three oriented films. Secondly, the PMA can be enlarged closely to $10^7$ erg/cm$^3$ ($7.34 \times 10^6$ erg/cm$^3$) in the [110]-oriented orthorhombic SRO. According the first-principle calculations, we find that the d-orbit band and MA of the Ru atom in the SRO are tightly affected by the orientation and crystal structure, leading the orientation- and structure-dependent MA. Finally, our films also exhibit reliable magnetism against bending test. These results demonstrate that the sacrificial layer is a powerful approach to fabricate high-quality flexible oxide films, and the structure and orientation are of importance to realize the strong and controllable PMA in the prepare membranes.

**RESULTS AND DISCUSSION**

Figure 1a shows the schematic process for fabricating flexible SRO films with different orientations, *i.e.* [001], [110] and [111]. SRO/SAO bilayers are deposited on the [001]-, [110]- and [111]-oriented $SrTiO_3$ substrates using pulsed laser deposition (PLD) (seen details in the METHODS). During the deposition, the thickness of the film is monitored by the reflection high-energy electron diffraction (RHEED) and controlled with 30 unit cells (u.c.) and 40 u.c. for the SAO and SRO, respectively. Then, the bilayers are immersed in pure water to etch the sacrificial SAO layers and separate the SRO membranes. During the etching process, flexible PDMS is used as a support layer to mechanically stabilize the films.[27] After releasing, the SRO/PDMS structures are transferred on desired rigid or flexible substrates. Then the freestanding membranes remain on the new substrates after peeling off the PDMS via thermally releasing. Finally, we obtain millimetre-scale and complete flexible SRO films as shown in Figure 1b. The surface and crystallization quality of the freestanding membranes are characterized by RHEED, atomic force microscopy (AFM), high-resolution transmission electron microscopy (HR-TEM) and selected-area electron diffraction (SAED) as shown in Figure 1c-e and Figure S1. These measurements show the SRO membranes are smooth with sub-nanometer roughness and high crystallization quality.

To explore the sacrificial layer influence on the magnetic properties of freestanding SRO membranes, we utilize different $Ca_{3-x}Sr_xAl_2O_6$ as sacrificial layers. In Figure 2a, we compare lattice parameters of various $Ca_{3-x}Sr_xAl_2O_6$ sacrificial layers with bulk SRO and STO. We choose $Ca_{1.5}Sr_{1.5}Al_2O_6$ (x= 1.5, CSAO) with a lattice constant $a_{CSAO}$= 4×3.87 Å, smaller than SRO and STO, while SAO has a larger lattice constant ($a_{SAO}$= 4×3.96 Å) than SRO and STO.[27, 37, 40] These different sacrificial layers can impose dissimilar strain in the SRO films. Here, we denote the SRO grown with SAO and CSAO as $SRO_S$ and $SRO_C$, respectively. Figure S2 and Figure S3 show smooth surface and high crystalline quality of the $SRO_C$, respectively, indicating the high quality of the freestanding $SRO_C$ (F-$SRO_C$) as observed in the freestanding $SRO_S$ (F-$SRO_S$). The X-ray diffraction (XRD) scans in Figure 2b and Figure S4 only show the peaks of the SRO, SAO/CSAO and STO substrate, revealing epitaxial growth and single phase of the as-grown (strained) and released (freestanding) SRO films.

More importantly, the systematical XRD measurements confirm the stress-strain related to the sacrificial layers. According to the 2$\theta$-$\omega$ scan curves, we calculate the interplanar distance ($d$) using the Bragg formula. As shown in Figure 2c, for the [001]-case, the $d$ of the strained $SRO_S$ (S-$SRO_S$) is 3.918 Å which is smaller than that of bulk pseudocubic SRO[18, 41] ($d$= 3.93 Å, marked with a green line), indicating that the SRO undergoes a tensile strain induced by the large lattice constant of the SAO. In contrast, the strained $SRO_C$ (S-$SRO_C$) grown with CSAO is compressed and its $d$ is elongated out-of-plane to 3.95 Å. These observations are also observed in the [110]- and [111]-oriented SRO films as shown in Figure 2c and Figure S4, demonstrating that the SRO films are tensely and compressively strained when grown with the SAO and CSAO sacrificial layer, respectively. Next, we carry out the XRD scans of the freestanding SRO (F-SRO) films attached on the PDMS. Figure 2b demonstrates that the peaks of SRO membranes shift to left (right) after dissolving the SAO (CSAO) sacrificial layers, further confirming the tensile (compressive) strain in the as-grown SRO$_S$ (SRO$_C$) films, respectively. The interplanar distances $d$ of the SRO$_S$ and SRO$_C$ films increase and decrease to 3.926 Å and 3.929 Å after lift-off, respectively. Correspondingly, the similar changes of the $d$ are also observed in [110] and [111] oriented membranes, as summarized in Figure 3c. Obviously, there is a stress release in the freestanding SRO films. However, it should be noted that the F-SRO$_S$ membranes have smaller $d$ compared to the F-SRO$_C$ membranes regardless of the crystal orientation. Such a difference may suggest a structural transition in SRO films induced by different sacrificial layers.

To verify the structural transition, we perform X-ray reciprocal space mapping measurements. As shown in Figure 2d, the mapping around (103)- and (013)-diffraction peaks of the SRO grown on STO without sacrificial layers reveal that (103)-peak of the SRO possess higher a $Q_z$ value than the (013)-peak ($\Delta Q_z \neq 0$). This indicates that the SRO is orthorhombic with a monoclinic distortion, in consistent with previous literatures.[18, 42-45] Similar results are observed for the strained and freestanding SRO films grown with the CSAO, in which different $Q_z$ values for the (103)- and (013)-diffraction peaks are observed (Figure S5b and Figure 2f). This implies S-SRO$_C$ and F-SRO$_C$ possess orthorhombic structure too. However, by comparing the peaks position for the films grown with the SAO (Figure S5a and Figure 2e), it is found that the (103)- and (013)-peaks have same $Q_z$ values ($\Delta Q_z = 0$), indicating that the S-SRO$_S$ and F-SRO$_S$ are tetragonal.[18, 42-45] These results demonstrate that the structural transition can be engineered via altering the sacrificial layers, as shown in Figure 2a. It is consistent with the literatures, in which the SRO films can form orthorhombic or tetragonal phases under the stress imposed by the substrate[18, 45], as we observed in the SAO/STO and CSAO/STO heterostructures. Both phases are almost degenerate in energy.[46, 47] On the other hand, these two crystal structures are belonging to different space groups that prohibit direct transformation among each other.[47] Therefore, these different crystal structures can be stabilized by the strain imposed by the sacrificial layers, and the structural phases could be kept even being peeled off from the substrates, leading the structural transition observed above.

The crystal orientations and observed structural transition have a remarkable influence toward the magnetic properties of flexible SRO films. To explore the impacts, magnetic hysteresis ($M$-$\mu_0H$) measurements were performed with the magnetic field in and out of the sample plane. As shown in Figure 3, the magnetization is more easily reversed and saturated in [110] F-SRO$_C$ membranes (Figure 3a-c) among the three orthorhombic samples, while it is easier in [001] tetragonal F-SRO$_S$ membranes (Figure 3e-f). Moreover, the SRO membranes exhibit very large MAEs above $1\times10^6$ erg/cm$^3$ (Figure 3d,h), confirming a robust MCA in SRO due to its strong SOC.[48] Intriguingly, the impacts of crystal orientations and structure on the MA are more pronounced. As shown in Figure 3a-c, the orthorhombic F-SRO$_C$ membrane, obtained with the CSAO sacrificial layer, shows an orientation-dependent PMA. The [110]-oriented F-SRO$_C$ membrane owns the largest anisotropy energy (MAE, $K_u$) of $7.34\times10^6$ erg/cm$^3$, which is 4.68 and 3.31 times of the films grown along [001] ($1.57\times10^6$ erg/cm$^3$) and [111] ($2.22\times10^6$ erg/cm$^3$) directions, as summarized in Figure 3d. In contrast, while in the tetragonal F-SRO$_S$ membrane grown with the SAO, not only the MAE is orientation-dependent, but also the magnetic easy-axis is controlled by the orientation (Figure 3e-h). At first, the $|K_u|$ of the [110]-film is the largest of $6.05\times10^6$ erg/cm$^3$ which is 1.9 and 1.34 times of the [001]- and [111]-films. Moreover, by altering the orientation from [001] to [110] and [111], the easy-axis rotates from out-of-plane to in-plane direction. These results demonstrate that the MA of the freestanding SRO film is not only dependent on the crystal orientation, but also the crystal structure which can be controlled further by the sacrificial layers.

To understand cooperative effects of crystal orientation and structure on the MA of freestanding SRO membranes, we perform first-principle density functional theory (DFT) calculations of the SRO MAE with the orthorhombic and tetragonal structure, as illustrated in Figure 2a (see details in METHODS). The calculated MAE is defined as the energy difference between the magnetization along the out-of-plane and in-plane. The positive and negative values of MAE represent perpendicular and in-plane MA, respectively. According to Figure 4a,d, firstly, the total energy calculation indicates that the SRO has a pronounced MCA with a high energy magnitude around ~1 meV/u.c. (~$10^7$ erg/cm$^3$). Secondly, both of the [110] orthorhombic and tetragonal phases have the largest MAE among three different crystal orientations. More importantly, the magnetic easy-axis can been tuned by the cooperative effect of the crystal orientation and structure. For the orthorhombic SRO, it shows a MA with a vertical easy-axis, exhibiting PMA for all three orientations. While for the tetragonal structure, the easy-axis is flipped from the out-of-plane direction to the plane when changing the orientation from [001] to [110] and [111]. These results are well consistent with our experimental observations, demonstrating that the cooperative effect of crystal orientation and structure in modulating SRO, not only the magnitude of magnetic anisotropy energy but also the direction of magnetic easy-axis.

To gain further insight into the cooperative modulation effect, we have performed element- and orbital-projected MAE calculations for both orthorhombic and tetragonal SRO based on a second-order perturbation theory (see details in METHODS). The element-resolved calculations show large MAEs (~1meV/u.c.) of the Ru atoms (Ru-MAEs) in both structural phases, as shown in Figure S7. This indicates that SRO possesses a large single ion anisotropy due to the strong spin-orbit interactions, in consistent with results of the SrIrO$_3$/La$_{0.67}$Sr$_{0.33}$MnO$_3$ heterostructures.[49] Figures 4b,e show the corresponding orbital-projected Ru-MAEs for the [110]-oriented orthorhombic and tetragonal SRO, respectively. In the orthorhombic phase, the ($d_{xz}$, $d_{yz}$) matrix element makes an outstanding out-of-plane contribution to the Ru-MAE, leading a PMA (Figure 4b) with a MAE of 0.937 meV/u.c.. However, for the tetragonal SRO, although the ($d_{xz}$, $d_{yz}$) keeps positive, most of the elements become negatively inducing an IMA with a MAE of -1.207 meV/u.c. (Figure 4e). It confirms that the tunable MA of SRO membranes are indeed controlled by Ru-MAEs through crystal orientation and structure. First-principle DFT calculations show the partial density of states (PDOS) around the Fermi level are solely derived from Ru atoms as shown in Figures 4c,f. In SRO, there is four electrons occupied the $t_{2g}$ orbitals due to the crystal field splitting. As shown in the inset of Figure 4c, in the orthorhombic structure, the $t_{2g}$ orbitals are almost three-fold degenerate. The degeneracy is partially lifted, however, in the tetragonal SRO. Two electrons are preferred occupied the $d_{xy}$ orbitals first, and the other electrons are located on the degenerate $d_{xz}$/$d_{yz}$ orbitals as illuminated in the inset of the Figure 4f. These crystal structure-dependent orbital splitting directly influence the PDOS at the Fermi level, which in turn affect the Ru-MAEs.

For actual application, the flexibility of the freestanding films should satisfy the reliability demands of flexible devices.[50] To probe the reliability, a bending cycle test is taken on the [111]-oriented orthorhombic SRO sandwiched between two 50-μm PDMS films and adhered on a PI tape, forming a structure of PDMS/SRO/PDMS/PI. For the test, as shown in Figure 5a, the testing structure is fixed by a mold which compresses the film with a radius of 2 mm, and then flattens without tensile stress as one bending cycle. Next, the magnetization dependent field and temperature ($M$-$\mu_0H$ and $M$-$T$) are characterized as a criterion for the endurance. Figure 5b shows the $M$-$\mu_0H$ curves after various bending cycles. During the testing, there is negligible change in the saturated and remanent magnetizations ($M_s$ and $M_r$) as well as the Curie temperature ($T_C$), indicating that the flexible membranes possess stable magnetic properties even after 10,000 bending cycles. We note that the coercive field ($H_C$) of flexible SRO membranes becomes enlarged after first 10 bending cycles and stabilizes in the following bending cycles. It may be owed to the cracks of the films generated during the bending testing.

**CONCLUSION**

In summary, we have successfully fabricated high-quality flexible SRO membranes with a strong and tunable PMA using the method of sacrificial layers. It is observed that the PMA, including MAE and magnetic easy-axis of the films, can be manipulated cooperatively by changing the growth orientation and sacrificial buffers, i.e. CSAO and SAO. The reason is that the crystal structure of the SRO film can be controlled by the buffers, and with the collaboration of the orientation. As a result, by optimizing the orientation along [110], we achieve a flexible PMA film with a maximum MAE of $7.34\times10^6$ erg/cm$^3$ in orthorhombic SRO when grown with the CSAO. The first-principle calculations indicate that the d-orbit band and MA of the Ru atom in the SRO is tightly affected by the orientation and crystal structure, leading the orientation- and structure-dependent MA. In addition, by performing a bending cycle test, the flexible membranes also show a good magnetism reliability. Our work exemplifies a feasible route to synthesis high-quality flexible PMA films using the method of sacrificial layer, and modify the PMA of the membranes by the collaboration of the structure and orientation. This strategy can be extended to a wide class of the oxide systems, and used to fabricate flexible PMA devices in future.

**METHODS**

*Film preparation.* The SAO/SRO and CSAO/SRO bilayers are deposited on different oriented STO substrates by pulsed laser deposition (PLD) using a KrF ($\lambda$=248 nm) excimer laser. The SAO layer with a thickness of 30 u.c. is deposited on the [001]-STO substrate at a substrate temperature $T_{sub}$=700 °C with an oxygen pressure $P_{O_2}$=2×10$^{-3}$ mbar. The laser fluence is 2.0 J/cm$^2$, and the repetition rate is 4 Hz. Subsequently, 40 u.c. SRO film is grown with the same laser flux at $T_{sub}$=700 °C and $P_{O_2}$=0.1 mbar. With same pulse counts and preparation condition of the SAO and SRO, the [110]- and [111]-bilayers are prepared. With same methods, the oriented CSAO/SRO bilayers are fabricated. Here, the CSAO layers are deposited with same condition as the SAO.

*Film transfer:* The films are adhered on clean polydimethylsiloxane (PDMS), then immersed in pure water at room temperature for 12 h to dissolve the SAO and CSAO buffers completely. Then, the SRO/PSMS structures are lifted out and washed clean with acetone. Finally, we obtain the different oriented flexible SRO films adhered on the PDMS. It should be noted that the SRO films can be transferred on other substrates or supports, such as the Si, STO or TEM grids after being baked at 90 °C for 10 min.

*Characterizations:* The information on the surface morphology, epitaxy quality and crystal structure are examined by atomic force microscopy (AFM, Dimension 3100, Bruker) and X-ray diffraction (XRD) scan (D8 Discover, Bruker). The high-resolution transmission electron microscopy and selected-area electron diffraction (HE-TEM and SAED, Talos F200CX, Thermal Fischer Scientific) are used to further characterize the crystalline quality. To perform the HE-TEM and SAED measurements for the flexible films, the freestanding SRO films are transferred on the TEM grids with holey carbon film. The magnetic properties are measured using superconducting quantum interference device (SQUID-VSM, Quantum Design) at 10 K. To measure the magnetism of the flexible SRO, the films are transferred on the clean STO substrates.

*First-principles calculation method:* First principles calculations were performed within the Vienna ab initio simulation package (VASP) code.[51-53] In all the calculations, the exchange-correlation potential was treated with the generalized gradient approximation (GGA) with the Perdew-Burke-Ernzerhof (PBE) functional.[54] The strong correlation effects for the Ru 4d is taken into account by means of the GGA+U approach and the on-site effective Coulomb interaction parameter $U_{eff}$ is taken as 2.0 eV.[55] The cutoff energy was 500 eV and 8×8×6 k-mesh was used in all simulations. To simulate the epitaxy, the horizontal lattice constants were fixed to those of the substrate, while the vertical one has been optimized until the force on each atom is smaller than 0.01 eV/Å. Furthermore, the magnetic anisotropy energy is calculated by using the force theorem.[7]

**ASSOCIATED CONTENT**

**Supporting Information**

Additional details on the surface morphology, epitaxy quality, crystal structure and magnetism properties of the as-grown and freestanding SRO films, element- and orbital-projected magnetic anisotropy energy (MAE) calculations for both orthorhombic and tetragonal SRO based on a second-order perturbation theory.

**AUTHOR INFORMATION**

**Corresponding Author**

**Xianfeng Hao** - *Key Laboratory of Applied Chemistry, College of Environmental and Chemical Engineering, Yanshan University, Qinhuangdao 066004, China;* E-mail: xfhao@ysu.edu.cn

**Zhiming Wang** - *CAS Key Laboratory of Magnetic Materials and Devices, Ningbo Institute of Materials Technology and Engineering, Chinese Academy of Sciences, Ningbo 315201, China; Zhejiang Province Key Laboratory of Magnetic Materials and Application Technology, Ningbo Institute of Materials Technology and Engineering, Chinese Academy of Sciences, Ningbo 315201, China; Center of Materials Science and Optoelectronics Engineering, University of Chinese Academy of Sciences, Beijing 100049, China;* E-mail: zhiming.wang@nimte.ac.cn

**Authors**

**Zengxing Lu** - *CAS Key Laboratory of Magnetic Materials and Devices, Ningbo Institute of Materials Technology and Engineering, Chinese Academy of Sciences, Ningbo 315201, China; Zhejiang Province Key Laboratory of Magnetic Materials and Application Technology, Ningbo Institute of Materials Technology and Engineering, Chinese Academy of Sciences, Ningbo 315201, China*

**Yongjie Yang** - *CAS Key Laboratory of Magnetic Materials and Devices, Ningbo Institute of Materials Technology and Engineering, Chinese Academy of Sciences, Ningbo 315201, China; Zhejiang Province Key Laboratory of Magnetic Materials and Application Technology, Ningbo Institute of Materials Technology and Engineering, Chinese Academy of Sciences, Ningbo 315201, China; Surface Engineering Institute, University of Science and Technology Liaoning, Anshan 114051, China*

**Lijie Wen** - *CAS Key Laboratory of Magnetic Materials and Devices, Ningbo Institute of*

Materials Technology and Engineering, Chinese Academy of Sciences, Ningbo 315201, China; Zhejiang Province Key Laboratory of Magnetic Materials and Application Technology, Ningbo Institute of Materials Technology and Engineering, Chinese Academy of Sciences, Ningbo 315201, China; Key Laboratory of Applied Chemistry, College of Environmental and Chemical Engineering, Yanshan University, Qinhuangdao 066004, China

**Jiatai Feng** - *CAS Key Laboratory of Magnetic Materials and Devices, Ningbo Institute of Materials Technology and Engineering, Chinese Academy of Sciences, Ningbo 315201, China; Zhejiang Province Key Laboratory of Magnetic Materials and Application Technology, Ningbo Institute of Materials Technology and Engineering, Chinese Academy of Sciences, Ningbo 315201, China*

**Bing Lao** - *CAS Key Laboratory of Magnetic Materials and Devices, Ningbo Institute of Materials Technology and Engineering, Chinese Academy of Sciences, Ningbo 315201, China; Zhejiang Province Key Laboratory of Magnetic Materials and Application Technology, Ningbo Institute of Materials Technology and Engineering, Chinese Academy of Sciences, Ningbo 315201, China*

**Xuan Zheng** - *CAS Key Laboratory of Magnetic Materials and Devices, Ningbo Institute of Materials Technology and Engineering, Chinese Academy of Sciences, Ningbo 315201, China; Zhejiang Province Key Laboratory of Magnetic Materials and Application Technology, Ningbo Institute of Materials Technology and Engineering, Chinese Academy of Sciences, Ningbo 315201, China; University of Nottingham, Ningbo 315201, China*

**Sheng Li** - *CAS Key Laboratory of Magnetic Materials and Devices, Ningbo Institute of Materials Technology and Engineering, Chinese Academy of Sciences, Ningbo 315201, China; Zhejiang Province Key Laboratory of Magnetic Materials and Application Technology, Ningbo Institute of Materials Technology and Engineering, Chinese Academy of Sciences, Ningbo 315201, China*

**Kenan Zhao** - *CAS Key Laboratory of Magnetic Materials and Devices, Ningbo Institute of Materials Technology and Engineering, Chinese Academy of Sciences, Ningbo 315201, China; Zhejiang Province Key Laboratory of Magnetic Materials and Application Technology, Ningbo Institute of Materials Technology and Engineering, Chinese Academy of Sciences, Ningbo 315201, China*

**Binshan Cao** - *CAS Key Laboratory of Magnetic Materials and Devices, Ningbo Institute of Materials Technology and Engineering, Chinese Academy of Sciences, Ningbo 315201, China; Zhejiang Province Key Laboratory of Magnetic Materials and Application Technology, Ningbo Institute of Materials Technology and Engineering, Chinese Academy of Sciences, Ningbo 315201, China; Nano Science and Technology Institute, University of Science and Technology of China, Hefei 230026, China*

**Zeliang Ren** - *CAS Key Laboratory of Magnetic Materials and Devices, Ningbo Institute of Materials Technology and Engineering, Chinese Academy of Sciences, Ningbo 315201, China; Zhejiang Province Key Laboratory of Magnetic Materials and Application Technology, Ningbo Institute of Materials Technology and Engineering, Chinese Academy of Sciences, Ningbo 315201, China; Nano Science and Technology Institute, University of Science and Technology of China, Hefei 230026, China*

**Dongsheng Song** - *Institutes of Physical Science and Information Technology, Anhui University, Hefei 230601, China*

**Haifeng Du** - *Anhui Key Laboratory of Condensed Matter Physics at Extreme Conditions, High Magnetic Field Laboratory, HFIPS, Anhui, Chinese Academy of Sciences, Hefei 230031, China.*

**Yuanyuan Guo** - *School of Materials and Metallurgy, University of Science and Technology Liaoning, Anshan 114051, China*

**Zhicheng Zhong** - *CAS Key Laboratory of Magnetic Materials and Devices, Ningbo Institute of Materials Technology and Engineering, Chinese Academy of Sciences, Ningbo 315201, China; Zhejiang Province Key Laboratory of Magnetic Materials and Application Technology, Ningbo Institute of Materials Technology and Engineering, Chinese Academy of Sciences, Ningbo 315201, China; Center of Materials Science and Optoelectronics Engineering, University of Chinese Academy of Sciences, Beijing 100049, China*

**Xianfeng Hao** - *Key Laboratory of Applied Chemistry, College of Environmental and Chemical Engineering, Yanshan University, Qinhuangdao 066004, China*

**Zhiming Wang** - *CAS Key Laboratory of Magnetic Materials and Devices, Ningbo Institute of Materials Technology and Engineering, Chinese Academy of Sciences, Ningbo 315201, China; Zhejiang Province Key Laboratory of Magnetic Materials and Application Technology, Ningbo*

*Institute of Materials Technology and Engineering, Chinese Academy of Sciences, Ningbo 315201, China; Center of Materials Science and Optoelectronics Engineering, University of Chinese Academy of Sciences, Beijing 100049, China*

**Run-Wei Li** - *CAS Key Laboratory of Magnetic Materials and Devices, Ningbo Institute of Materials Technology and Engineering, Chinese Academy of Sciences, Ningbo 315201, China; Zhejiang Province Key Laboratory of Magnetic Materials and Application Technology, Ningbo Institute of Materials Technology and Engineering, Chinese Academy of Sciences, Ningbo 315201, China; Center of Materials Science and Optoelectronics Engineering, University of Chinese Academy of Sciences, Beijing 100049, China*

**Notes**

The authors declare no competing financial interest.

**ACKNOWLEDGMENTS**

This work was supported by the National Key Research and Development Program of China (Nos. 2017YFA0303600, 2019YFA0307800), the National Natural Science Foundation of China (Nos. U1832102, 11874367, 51931011, 51902322), the Key Research Program of Frontier Sciences, Chinese Academy of Sciences (No. ZDBS-LY-SLH008), the Thousand Young Talents Program of China, the 3315 Program of Ningbo, the Natural Science Foundation of Zhejiang province of China (No. LR20A040001), the Public Welfare Technical Applied Research Project of Zhejiang Province (No. LY21E020007), the Ningbo Natural Science Foundation (No. 2019A610050), the Beijing National Laboratory for Condensed Matter Physics.

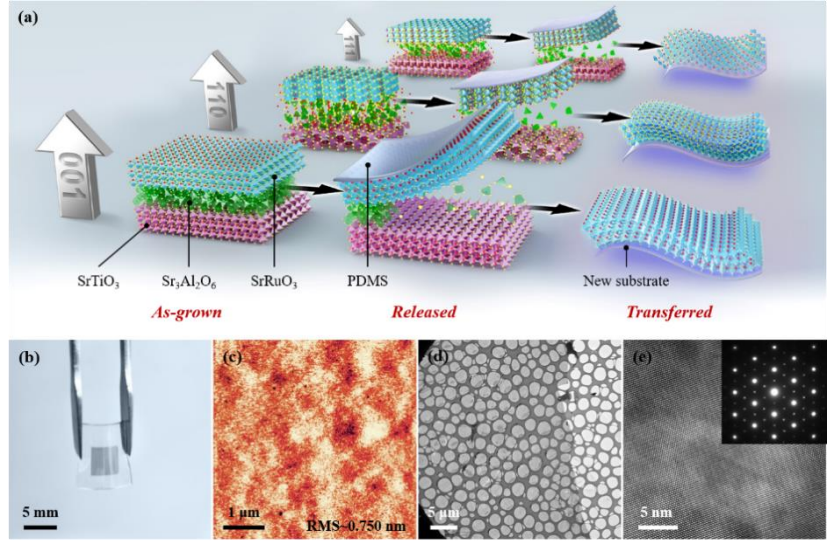

Figure 1. **Preparation of flexible SRO membranes with different crystal orientations.** (a) Schematic illustrating the heterostructure growth, lift-off and transfer process for SRO membranes with different orientations. During the preparation, the water-soluble $Sr_3Al_2O_6$ is used as sacrificial layers. (b) Flexible SRO membranes (gray area) transferred on soft PDMS. (c) An AFM image of the freestanding SRO membrane with a sub-nanometer roughness (RMS). (d) Low-magnification and (e) atomically resolved plane-viewed HR-TEM image of [111]-oriented freestanding SRO membrane transferred on a TEM grid with holey carbon film. The inset in (e) shows the SAED pattern.

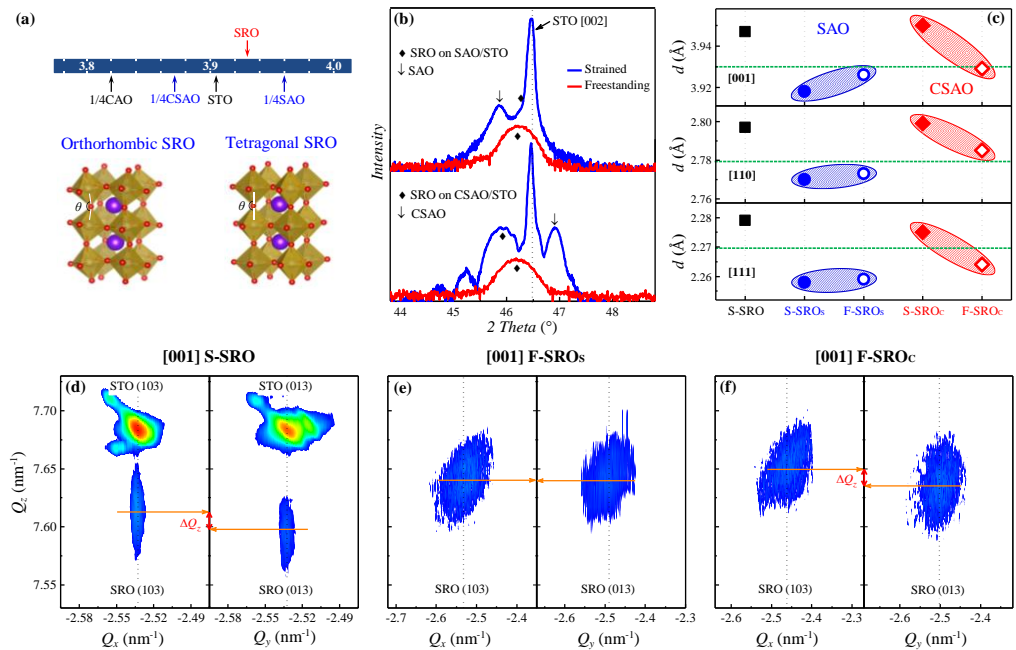

Figure 2. **Crystalline structure of epitaxial freestanding SRO films grown with different sacrificial layers.** (a) Comparative lattice constants (Å) of the pseudocubic SRO, sacrificial layers (i.e. SAO and CSAO) and STO substrate. (b) XRD $2\theta$-$\omega$ scans of the strained and freestanding SRO films with the SAO (TOP) and CSAO (bottom). (c) Interplanar distance $d$ of the strained and freestanding [001]-, [110]- and [111]-oriented SRO films grown on STO, SAO/STO (blue) and CSAO/STO (red).

Reciprocal space maps around the (103) and (013) peaks for the (d) [001]-oriented SRO films grown on STO substrate, [001]-oriented freestanding SRO films grown with (e) SAO and (b) CSAO. S-SRO, S-SRO$_S$ and S-SRO$_C$ represent the stained SRO, SRO grown with SAO and CSAO films. F-SRO$_S$ and F-SRO$_C$ represent the freestanding SRO grown with SAO and CSAO.

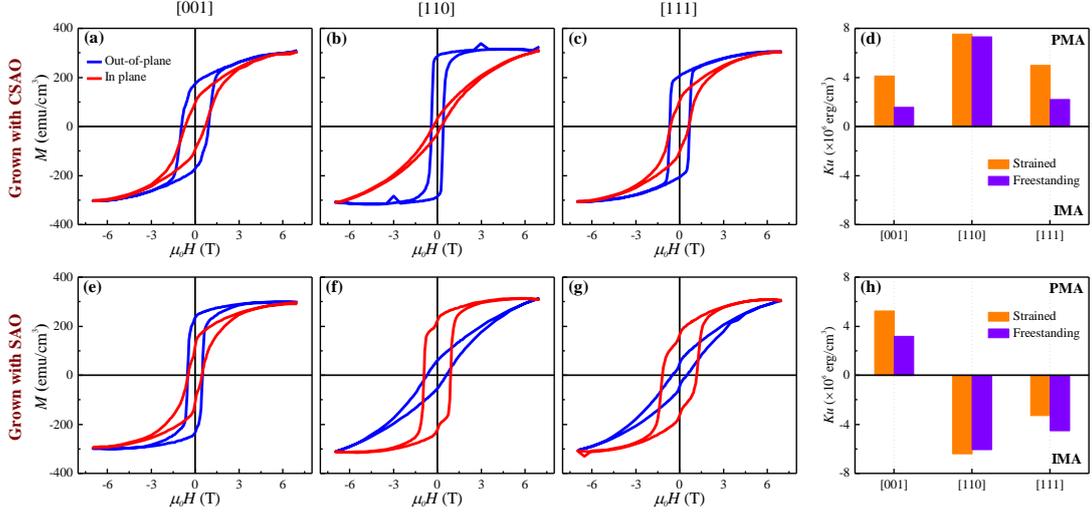

Figure 3. **MA of the freestanding SRO films modulated by the crystal orientation and structure**. MA of the (a) and (e) [001]-, (b) and (f) [110]-, (c) and (g) [111]-oriented freestanding SRO films grown with the CSAO and SAO, respectively. CSAO: (a), (b) and (c). SAO: (e), (f) and (g). MAE dependent on the orientation in the freestanding SRO films grown with the (d) CSAO and (h) SAO, respectively. PMA and IMA in (d) and (h) are the abbreviations of perpendicular and in-plane magnetic anisotropy, respectively.

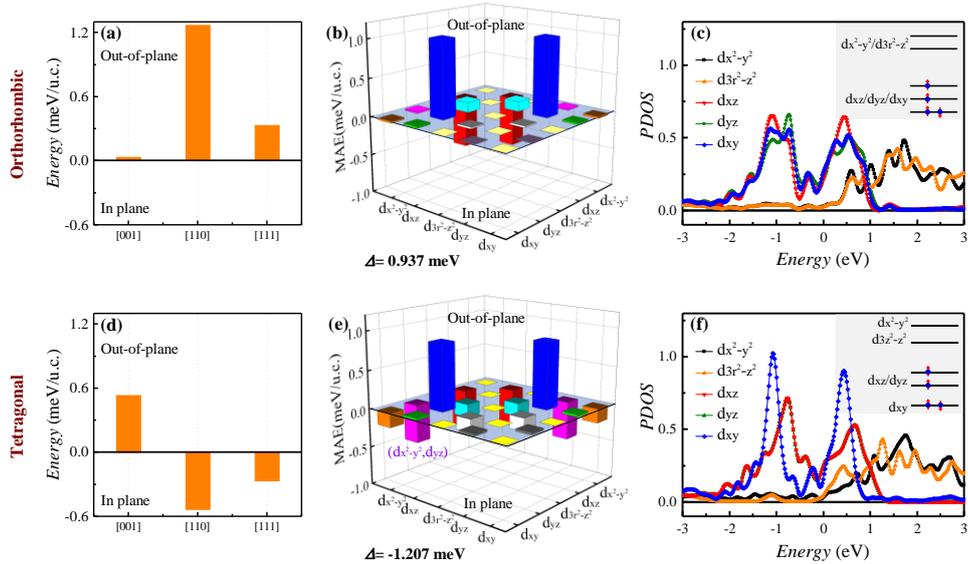

Figure 4. **Mechanism of the orientation- and structure-dependent MA.** The MAE of the (a) orthorhombic and (d) tetragonal SRO dependent on the orientation, respectively. The d orbit-resolved MAE of the Ru atoms in [110]-oriented (b) orthorhombic and (e) tetragonal SRO. The d orbit-resolved PDOSs of the Ru atoms in (c) orthorhombic and (f) tetragonal SRO. The insets in (c) and (f) shows the d-orbit energy band structures modulated by the structure.

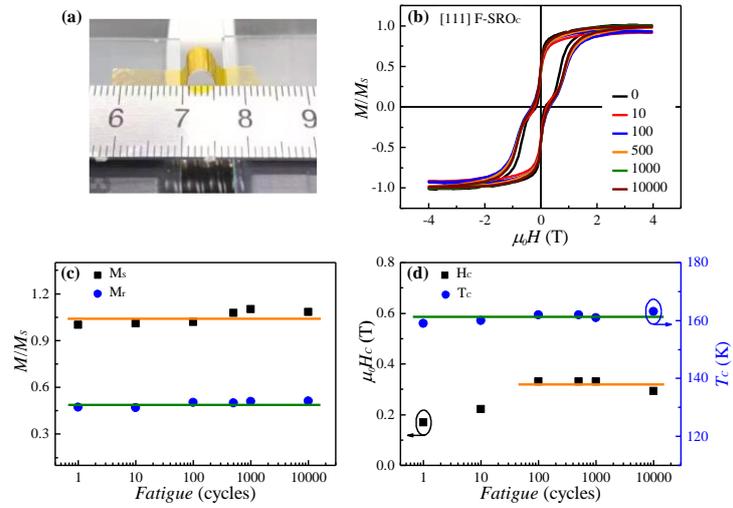

Figure 5. **The reliability of the flexible SRO films.** (a) Sketch map of the bending cycle test. (b) The normalized $M$-$\mu_0H$ loops of films in flat state measured after different bending cycles. (c) The plot of saturated and remanent magnetizations ($M_s$ and $M_r$) versus bending cycles. (d) The plot of Curie temperature and coercive field ($H_C$ and $T_C$) versus bending cycles. During the test, the [111]-oriented orthorhombic SRO film, sandwiched between two 50-μm PDMS films and adhered on a PI tape, is used and bent to 2 mm radius.

# Supporting Information

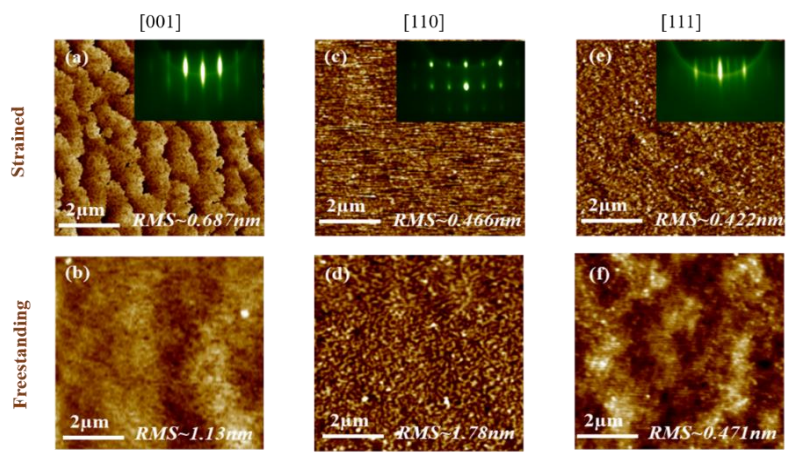

Figure S1. **Surface morphology of different oriented SRO films grown with the SAO before and after lift-off.** AFM images of SRO films with (a) and (b) (001), (c) and (d) (110), (e) and (f) (111) crystal orientations. The top and bottom images show the surfaces of the strained and freestanding films, respectively. Insets in the (a,c,e) show the RHEED patterns of the initial films grown with different crystalline orientations. RMS represents the roughness of the film.

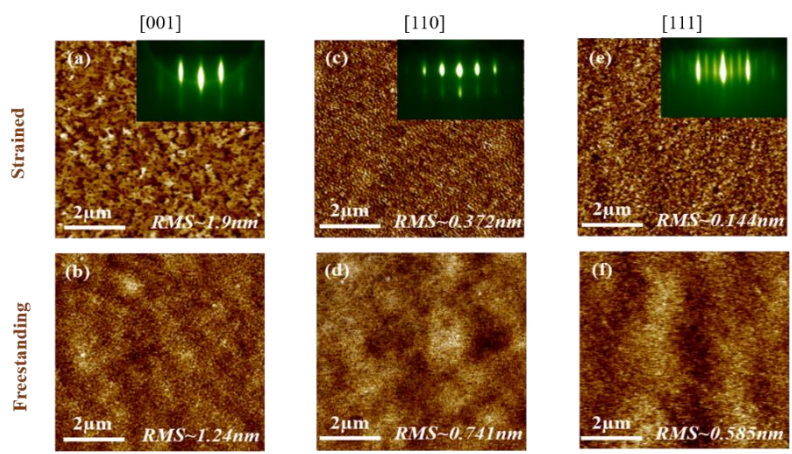

Figure S2. **Surface morphology of different oriented SRO films grown with the CSAO before and after lift-off.** AFM images of SRO films with (a) and (b) (001), (c) and (d) (110), (e) and (f) (111) crystal orientations. The top and bottom images show the surfaces of the strained and freestanding films, respectively. Insets in the (a,c,e) show the RHEED patterns of the initial films grown with different crystalline orientations. RMS represents the roughness of the film.

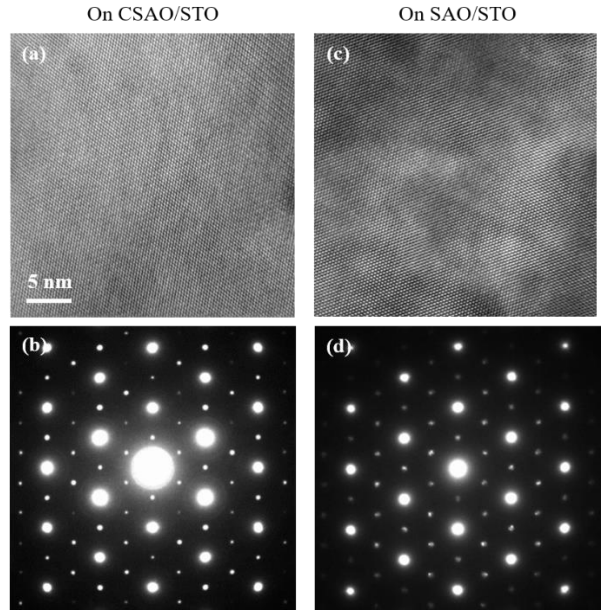

Figure S3. **Structural characterizations of the [111]-oriented freestanding SRO films.** Low-magnification plane-viewed HR-TEM image of [111]-oriented freestanding SRO membrane grown with the (a) CSAO and (c) SAO, respectively. (b) and (d) Atomically resolved plane-viewed SAED images of both films. For the measurement, the films are transferred on TEM grids with holey carbon film.

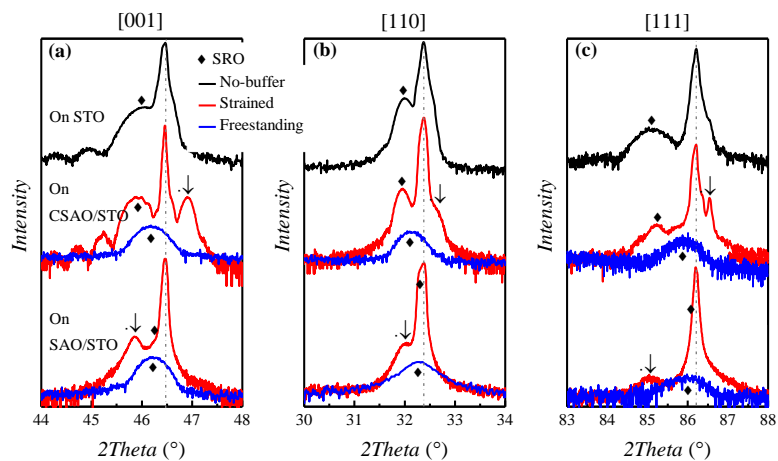

Figure S4. **Epitaxial growth of the SRO films.** XRD $2\theta$-$\omega$ scans of the (a) [001]-, (b) [110]- and (c) [111]-oriented strained and freestanding SRO films grown on STO, CSAO/STO and SAO/STO. Symbol ↓ represents the sacrificial layers.

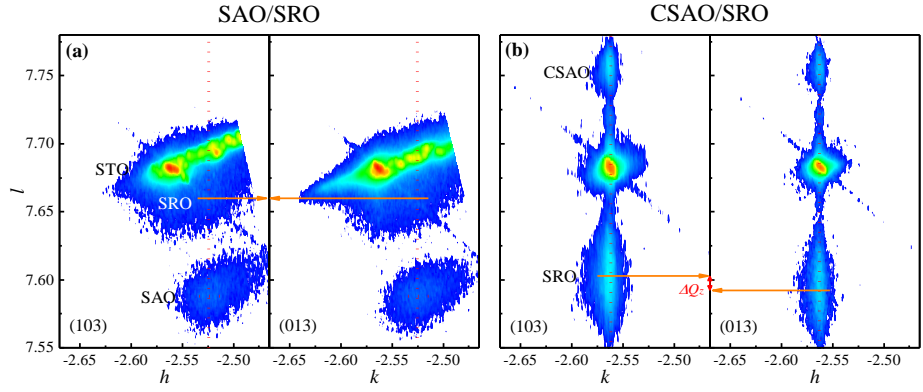

Figure S5. **Structural characterizations of the [001]-SRO films grown with SAO and CSAO.** RSMs around the (103) and (013) peaks for the films grown with (a) SAO and (b) CSAO.

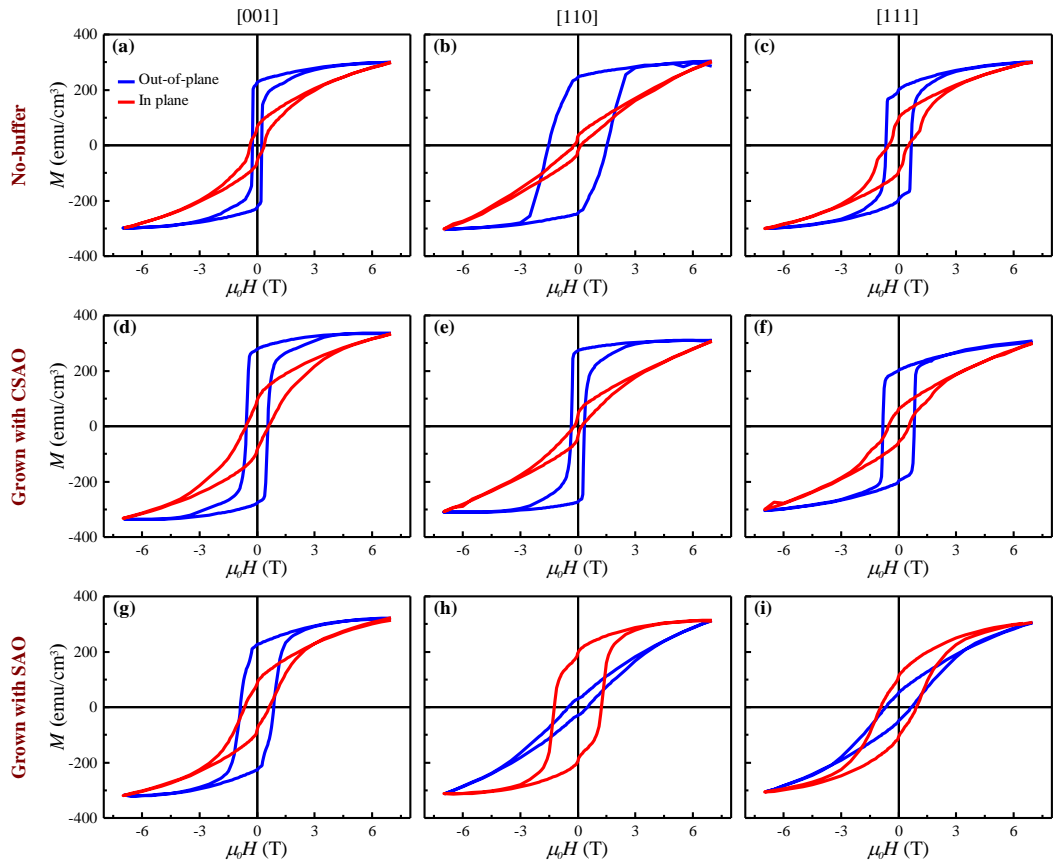

Figure S6. **MA of the as-grown SRO films modulated by the crystal orientation and structure.** MA of the (a,d,g) [001]-, (b,e,h) [110]- and (c,f,i) [111]-oriented SRO films grown on STO, CSAO/STO and SAO/STO.

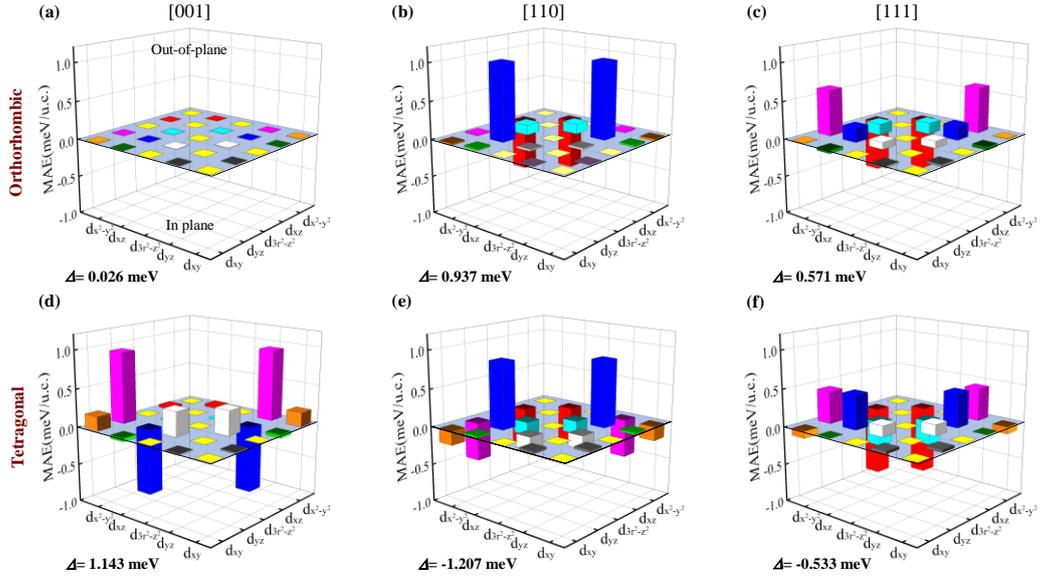

Figure S7. **Orientation- and structure-dependent MAE.** The *d* orbit-resolved MAE of the Ru atoms in the [001]-, [110]- and [111]-oriented (a)-(c) orthorhombic and (d)-(e) tetragonal SRO.